\documentclass[pra,twocolumn,showpacs,lengthcheck,preprintnumbers,longbibliography,superscriptaddress]{revtex4-1}
\usepackage{epsfig}
\usepackage{amsmath}
\usepackage{amsfonts}
\usepackage{fancyhdr}
\usepackage{amsmath}
\usepackage{amssymb}
\usepackage{mathtools}
\usepackage{graphicx}
\usepackage[unicode]{hyperref}
\hypersetup{
   unicode=true,          % non-Latin characters in AcrobatÕs bookmarks
   plainpages=false,
   colorlinks=true,       % false: boxed links; true: colored links
   citecolor=blue,        % color of links to bibliography
}
\newcommand{\node}{\mathmbox{\;\mid\mspace{-14.55mu}\times}} % assumes math mode
\begin{document}

\title{Nonlinear standing waves in an array of coherently coupled Bose-Einstein condensates}
\author{Christian Baals}
\affiliation{Department of Physics and Research Center OPTIMAS, Technische 
Universit\"at Kaiserslautern, 67663 Kaiserslautern, Germany}
\affiliation{Graduate School Materials Science in Mainz, Staudinger Weg 9, 55128 Mainz, Germany}
\author{Herwig Ott}
\affiliation{Department of Physics and Research Center OPTIMAS, Technische 
Universit\"at Kaiserslautern, 67663 Kaiserslautern, Germany}
\author{Joachim Brand}
\affiliation{Dodd-Walls Centre for Photonic and Quantum 
Technologies,  
Centre for Theoretical Chemistry and Physics, New Zealand 
Institute for Advanced Study, Massey University, Private Bag 102904 NSMC, 
Auckland 0745, New Zealand}
\author{Antonio Mu\~noz Mateo}
\affiliation{Dodd-Walls Centre for Photonic and Quantum 
Technologies,  
Centre for Theoretical Chemistry and Physics, New Zealand 
Institute for Advanced Study, Massey University, Private Bag 102904 NSMC, 
Auckland 0745, New Zealand}

\date{\today}

\begin{abstract}
Stationary solitary waves are studied in an array of $M$ linearly-coupled one-dimensional Bose-Einstein condensates (BECs) by means of the  Gross-Pitaevskii equation. Solitary wave solutions with the character of overlapping dark solitons, Josephson vortex -- antivortex arrays, and arrays of half-dark solitons are constructed for $M>2$ from known solutions for two coupled BECs. Additional solutions resembling vortex dipoles and rarefaction pulses are found numerically. 
Stability analysis of the solitary waves reveals that
overlapping dark solitons can become unstable and
susceptible to decay into arrays of Josephson vortices. The Josephson vortex arrays have mixed  stability but for all parameters we find at least one stationary solitary wave configuration that is dynamically stable. The different families of nonlinear standing waves bifurcate from one another. In particular we demonstrate that Josephson-vortex arrays bifurcate from dark soliton solutions at instability thresholds. The stability thresholds for dark soliton and 
Josephson-vortex type solutions are provided, suggesting the feasibility of  
realization with optical lattice experiments.
\end{abstract}

%\pacs{03.75.Lm, 67.85.De, 37.10.Jk}
\maketitle

\section{Introduction}

The Josephson effect is a prominent manifestation of a macroscopic 
quantum phenomenon. Since its prediction in the 1960s, it has been thoroughly studied in 
superconducting systems, and has found plenty of applications in 
electronic devices \cite{Barone}. The effect relies on the 
existence of a complex order parameter describing the dynamics of an underlying 
Bose condensate of Cooper pairs, and due to this fact the associated phenomenology was soon 
foreseen and later observed in superfluid helium 
\cite{Anderson1966,Khorana1969,Hoskinson2005}.
More recently, the advent of Bose-Einstein condensates (BECs) of ultracold 
gases has opened many new possibilities for the realization of the Josephson 
effect. The Josephson equations for the modulus and relative phase of the 
order parameter, accounting for the superfluid particle number density and the
superfluid velocity, respectively, predict a non-dissipative flow across a 
barrier separating two superfluids. Such a flow has been monitored in 
experimental realizations by direct observation of the particle current versus 
chemical potential \cite{Cataliotti2001,Albiez2005,Levy2007}, or even the 
phase-current relationship  \cite{Eckel2014}. These fundamental developments 
are also promising steps for advances in the emerging field of atomtronics 
\cite{Amico2017}.

Most of the research about the Josephson effect in ultracold gases has 
addressed the case of  pointlike junctions, based on the study case of a BEC in
a double well potential \cite{Giovanazzi2000}. 
Long Josephson junctions that allow for a varying 
phase along the junction between BECs \cite{Kaurov2005} have 
comparatively received much less attention, in spite of the fact that these 
systems can support a very interesting topological structure: the Josephson 
vortex, or fluxon \cite{Barone}. These topological objects involve 
localized supercurrents and have found technical
application in the field of superconductors, e.g. they could be used as information carrying
computational bits, or even qubits, due to the fact that they can trap magnetic flux  
\cite{Devoret2013,Roditchev2015}. 
In BECs, evidence for spontaneously formed Josephson vortices in linearly coupled one-dimensional BECs during a rapid cooling procedure has only been reported very recently \cite{Schweigler2017} (where they were identified as ``sine-Gordon solitons'') after such a mechanism had been proposed theoretically \cite{Su2013}.

Theoretical studies on Josephson vortices in BECs have mainly  
considered  systems with two linearly coupled 1D condensates
\cite{Kaurov2005,Kaurov2006,Brand2009,Qadir2012}, and a one-dimensional spinor 
BEC with internal Josephson effect \cite{Son2002,Qu2017}. 
The  models used  to describe coupled 1D BECs start from two linearly coupled Gross-Pitaevskii equations (GPEs) \cite{Kaurov2005}, or a coupled Luttinger-liquid model \cite{Gritsev2007}. In the small coupling limit, a sine-Gordon model for the relative phase of the two condensates can be derived where the Josephson vortex becomes a sine-Gordon soliton \cite{Kaurov2005,Gritsev2007,Sophie2017}.
Recently the full dispersion relation of moving Josephson vortices and related solitary excitations in the more general coupled Gross-Pitaevksii model has been found \cite{Sophie2017}. Closely related oscillon excitations, which can be understood as long-lived bound states of Josephson vortices were described in Ref.~\cite{Su2015}.
Generalizations to tunnel-coupled spinor gases 
\cite{Montgomery2013}, and multidimensional Bose gases \cite{Gallemi2016} have also 
been considered.
But there is an interesting system that has not yet been explored, namely 
the appearance of Josephson vortices in an array of more than two BECs. 
Such an arrangement could result from slicing a BEC with an optical lattice in the tight binding regime. The arrangement is analogous to layered superconductors \cite{Klemm} where Josephson junctions between layers are coupled, but the theoretical model used for coupled Josephson junctions in superconductors \cite{Kivshar1988} differs from the coupled Gross-Pitaevskii model used to describe  arrays of coupled BECs.

In this paper, we study  patterns of Josephson vortices and related solitary waves  in 
arrays of  linearly coupled 1D BECs.
The whole arrangement forms a discrete version of 
a single two-dimensional (2D) BEC that could be experimentally realized by subjecting a 2D BEC to a 1D optical 
lattice in the tight binding regime. Our starting point is the analytical 
solution of an array of identical, static dark solitons in the 1D BECs.
A dark soliton in a single 1D BEC is a  well known, 
stable nonlinear wave. The linear coupling in an array of  condensates breaks this 
condition and can eventually lead to the decay of the solitons. As we will show, 
however, the stability is (maybe counter-intuitively) preserved for high values 
of the coupling, whereas the decay into Josephson vortices takes place at low 
coupling. 
We characterize a number of possible stationary solitary-wave solutions along with their stability and classify their character in terms of Josephson vortices, solitons, and half-solitons by means of a symbolic representation.  Although not all possible stationary solitary waves are stable, there is at least one stable configuration that may remain as a final result of decay processes of unstable solitary waves. We also find situations of multi-stability, where several stable solitary-wave solutions coexist.

The situation of extended dark soliton and Josephson vortex excitations in a
stack of coupled 1D BECs is closely related to dark soliton stripes (or planar 
dark solitons) in 2D or 3D BECs. The snaking instability of the dark soliton, which leads to decay into vortex structures, was observed in BECs \cite{Anderson2001} and superfluid Fermi gases \cite{Ku2016} and studied theoretically  in Refs.\ \cite{Kuznetsov1988,Muryshev1999,Brand2002,Cetoli2013}. In 
close analogy to the results reported in this work, the instability thresholds 
give rise to bifurcations of symmetry-breaking vortex-type solutions 
\cite{Brand2002,Komineas2003,Mateo2015a,Mateo2014}, of which a single vortex 
line, also called solitonic vortex, is the only dynamically stable nonlinear 
wave \cite{Brand2002,Toikka2016}.

This paper is organized in the following way: Section \ref{sec:sys}  starts with introducing
the model (Sec.~\ref{sec:GPmodel}) and overlapping dark soliton solutions (Sec.~\ref{sec:dss}). After discussing the solitary-wave solutions known for two coupled BECs in Sec.~\ref{sec:twocoupled}, solutions for arrays with an even number of BECs are constructed from the known solutions for two BECs in Sec.~\ref{sec:rep}, and more general solutions are briefly discussed in Sec.~\ref{sec:other}.
Linear stability analysis of the stationary 
solutions is performed in Sec.~\ref{sec:Bog} for a ring configuration of coupled BECs, where analytic results for 
stability thresholds of the dark soliton stack are discussed as well as numerically obtained spectra of 
unstable modes. The bifurcation scenario of Josephson vortex solutions is discussed in 
Sec.~\ref{sec:JV} before concluding with a discussion of possible experimental realizations in Sec.~\ref{sec:concl}.

\section{System: stack of BECs} \label{sec:sys}

\begin{figure}[t!]
\includegraphics[width=8.5cm]{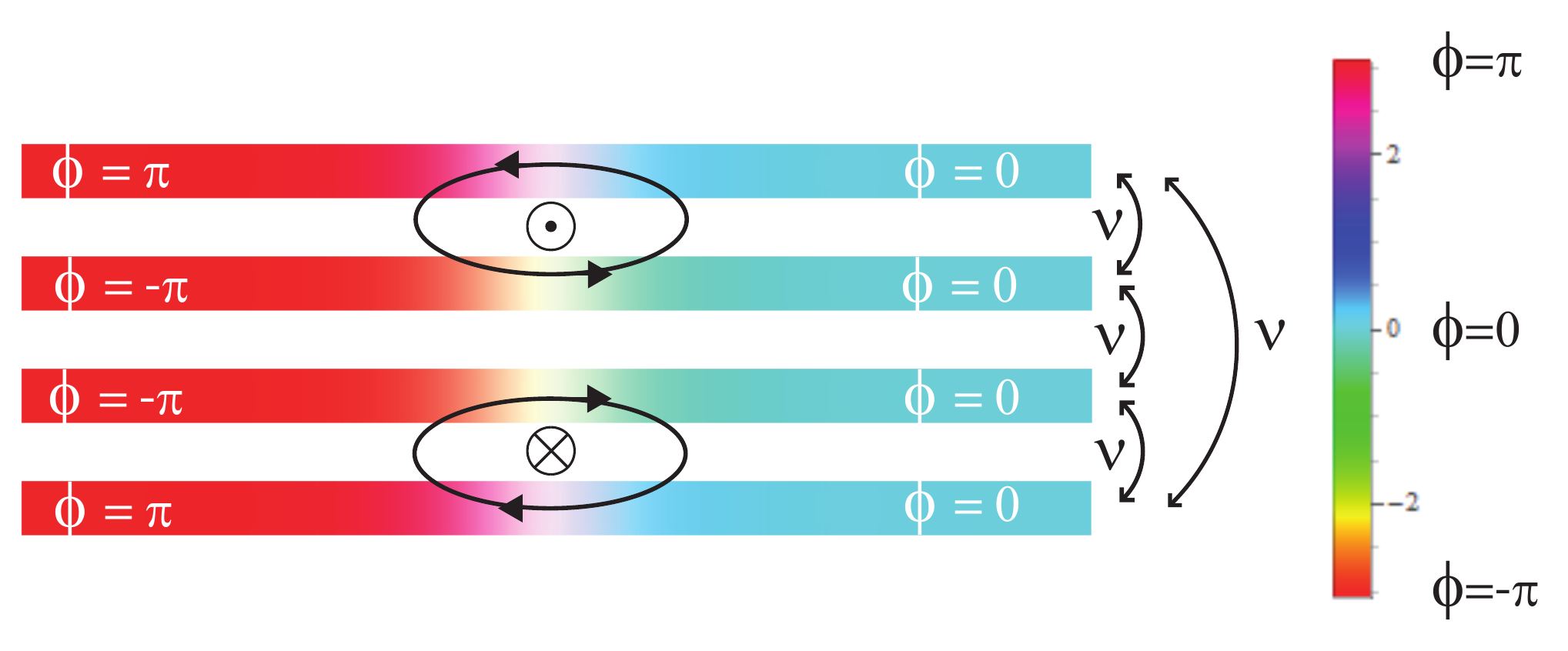}
\caption{Schematic of an array with $M=4$ coupled 1D BECs described by 
Eq.\ \eqref{eq:tgpe} with linear coupling $\nu$ effective between neighboring 
BECs in the ring configuration (the planar configuration has an identical solution). Shown is a stationary solitary-wave solution in the symmetric Josephson vortex -- anti-vortex configuration $a^*$. The color
encodes the complex phase (per color bar on right) and contrast encodes density. 
From the phase gradients (differences) encoding superfluid currents 
(Josephson currents) it can be deduced that the configuration contains one 
Josephson vortex and one anti-vortex, as indicated. Symbolically, we represent 
the configuration by  $\downarrow 
\odot\uparrow \:\uparrow \otimes \downarrow$.
}
\label{fig:schem}
\end{figure}  

\subsection{Gross-Pitaevskii model} \label{sec:GPmodel}

Consider an array of $M$ one dimensional BECs with linear coherent tunnel
coupling between nearest neighbors  arranged as shown in Fig.\ \ref{fig:schem}. 
Both the ring (periodic) and planar (open) configurations of the 
array can be modeled by wave functions 
$\Psi_j(x,t)$ with
$j=1,2,...M$,
satisfying the coupled GPEs 
\begin{equation}
i\hbar\frac{\partial\Psi_j}{\partial t}  = \frac{-\hbar^2}{2m} 
\partial_{xx}\Psi_j + g \left\vert \Psi_j\right\vert ^{2}\Psi_j + \epsilon_j \Psi_j
- \nu \,  
(\Psi_{j-1}+\Psi_{j+1}) \,,
\label{eq:tgpe}
\end{equation}
where $|\Psi_j(x,t)|^2 = n_j(x,t)$ represents the particle number density in
each component. The linear coupling energy between neighboring condensates is  
$\nu\ge 0$  and $\epsilon_j$ is an energy bias that could, e.g., encode a 
trapping potential. For simplicity, we assume that the  1D inter-particle 
interaction strength $g = 2\hbar \omega_\perp a$ is constant across the BEC 
array, where $a$ is the $s$-wave scattering 
length and we have assumed that the 1D regime has been reached within a 
tight transverse trap of frequency $\omega_\perp$. 
In order to simplify the theoretical analysis and allow for closed-form
analytical solutions, we assume the BECs to have infinite extent in $x$ 
direction and consider a specific choice for the energy bias  
$\epsilon_j$:\begin{itemize}
\item 
The \emph{planar configuration} uses  Dirichlet-type boundary conditions
obtained by setting  $\Psi_0 \equiv 0 \equiv \Psi_{M+1}$ and an energy bias of 
\begin{align}
\epsilon_1 &= \epsilon_M = -\nu\\
\epsilon_j &= 0 \quad \textrm{for} \quad 1<j<M.
\end{align}
\item
The \emph{ring configuration} uses periodic boundary conditions along the stack
where $\Psi_0 \equiv \Psi_M$ (and $\Psi_1 \equiv \Psi_{M+1}$) with vanishing 
energy bias  $\epsilon_j=0$. 
\end{itemize}
These conditions idealize the typical conditions that may be achieved in
experiments simply for the purpose of finding simple closed-form solutions of 
the GPEs.

The stationary states $\Psi_j(x,t) =\psi_j(x) e^{-i\mu t /\hbar}$ , with chemical 
potential $\mu$, satisfy the time-independent coupled GPEs
\begin{equation}
0 = \left(\frac{-\hbar^2}{2m} \partial_{xx}+ g|\psi_j|^2 -\mu+ \epsilon_j
\right)\psi_j -\nu \left(\psi_{j-1}+\psi_{j+1}\right) \,.
\label{eq:gpe}
\end{equation}
Stationary solutions corresponding to a constant density $n$ in each component
are found with $\psi_j(x) = \sqrt{n} \exp(i\Phi)$ with $g n=\mu+2 \nu>0$ and 
arbitrary constant global phase $\Phi$. 

For solitary waves solutions that only deviate locally from a constant
background we define the solitary wave energy $E_s$ as usual by 
\begin{align} \label{eq:Es}
E_s = \int \left[(\varepsilon_\mathrm{sol} - \mu n^\mathrm{tot}_\mathrm{sol}) -  (\varepsilon_\mathrm{bg} - \mu n^\mathrm{tot}_\mathrm{bg})\right] dx ,
\end{align}
where $\varepsilon_\mathrm{sol}$ and $\varepsilon_\mathrm{bg}$ are the Gross-Pitaevskii energy
densities of the solitary wave and background solutions, respectively, given by
\begin{align}
\nonumber
\varepsilon (x) =& \sum_{j=1}^M \frac{\hbar^2}{2m}\left|{\partial_x \psi_j}\right|^2 + \frac{g}{2} |\psi_j|^4
+\epsilon_j  |\psi_j|^2 \\
 &- \nu \psi_j^*(\psi_{j-1} +\psi_{j+1}) ,
\end{align}
and the total particle number density is given by 
\begin{align}
n^\mathrm{tot}(x) = \sum_{j=1}^M |\psi_j|^2 .
\end{align}
The solitary wave solutions are localized and exponentially heal against the
constant background, which is to be taken at the same chemical potential. 
Therefore, the integral in Eq.\ \eqref{eq:Es}, which formally runs over an 
infinite domain, converges and in practice can be taken over a finite interval 
that is large compared to the localization length scale of the solitary wave.
For the constant background, we find $n^\mathrm{tot}_\mathrm{bg} = M n$ and 
$\varepsilon_\mathrm{bg} = Mgn^2/2-2M\nu n$.

\subsection{Dark soliton stack}\label{sec:dss}

The stationary coupled GPEs (\ref{eq:gpe}) admit dark solitons solutions with identical wave functions in each BEC for both the planar and periodic configurations
\begin{equation}
 \psi_j(x)=\sqrt{n}\, \tanh\left(\frac{x}{\xi}\right) \,,
 \label{eq:dark}
\end{equation}
where $n = (\mu+2 \nu)/g$ is the constant background density and  the healing lengh $\xi =\hbar/\sqrt{mgn}$ sets the relevant length scale for the soliton. The solution \eqref{eq:dark} represents a kink in the wave function with a sign change at the identical position  (here $x=0$) in all coupled condensates and completely aligned phases. 
We refer to the solution \eqref{eq:dark} as the \emph{dark soliton stack}. 

Note that the dark soliton stack solution \eqref{eq:dark} is independent of the linear coupling $\nu$ when written in terms of the background density $n$ and $g$.  The solution has vanishing mass current density $j_j(x)$  in each component, defined by
\begin{align}
\nonumber
j_j(x) &= \frac{\hbar}{2 i m}\left(\Psi^*_j \partial_x \Psi_j - \Psi_j \partial_x \Psi^*_j\right)\\
 &= n_j(x)  \frac{\hbar}{m}\partial_x \arg\Psi_j ,
\end{align}
as well as vanishing Josephson current density $j_{j,j+1}(x) \propto \arg \Psi_{j+1}-\arg \Psi_j$. We symbolically denote the dark soliton stack with lined-up kinks in the BECs by $\node\node\node\node$
for the case of $M=4$ BECs, as opposed to the node- and current-less background (ground-state) denoted by $\mid\;\mid\;\mid\;\mid$. 

The energy of the dark soliton stack is $E_s^\mathrm{DSS} = M E_\mathrm{DS}$, where 
\begin{align}
E_\mathrm{DS} = \frac{4}{3} \frac{\sqrt{g}n^\frac{3}{2}\hbar}{\sqrt{m}}
\end{align}
is the energy of a dark soliton in a single (scalar) one-dimensional BEC, as is easily verified from Eq.\ \eqref{eq:Es}.

\subsection{Two coupled BECs}\label{sec:twocoupled}

Let us first consider the case of $M=2$ coupled BECs, which has already been studied extensively \cite{Kaurov2005,Qadir2012, Sophie2017}. Let us write the stationary coupled GPEs explicitly
\begin{align}
\nonumber
0 &= \left(\frac{-\hbar^2}{2m} \partial_{xx}+ g|\psi_1|^2 -\tilde{\mu}
\right)\psi_1 -\tilde{\nu}\psi_{2},\\
0 &= \left(\frac{-\hbar^2}{2m} \partial_{xx}+ g|\psi_2|^2 -\tilde{\mu}
\right)\psi_2 -\tilde{\nu} \psi_{1} .
\label{eq:gpe2}
\end{align}
The planar and ring configuration become identical in this case of  $M=2$ as the form \eqref{eq:gpe2}
is obtained from the more general GPEs \eqref{eq:gpe} in both the planar configuration, where $\tilde{\nu}=\nu$ and $\tilde{\mu} = \mu - \nu$, and in the ring configuration, where $\tilde{\nu}=\nu/2$ and  $\tilde{\mu} = \mu$.

The constant density solutions with $\tilde{\mu} +\tilde{\nu} = gn$ provide the background for the dark soliton stack \eqref{eq:dark}, which exists for any value of $\tilde{\nu}\geq 0$. In addition,
stationary \emph{Josephson vortex} solutions exist when $\tilde{\nu}\leq gn/4$  and take the form \cite{Kaurov2005}
\begin{align}
\nonumber
 \psi_{1}^{(\otimes)}(x)&=\sqrt{n}\, \left[\tanh\left(\frac{x}{\xi_{\tilde{\nu}}}\right)+ i 
\sqrt{1-\frac{4\tilde{\nu}}{gn}}\,\mbox{sech}\left(\frac{x}{\xi_{\tilde{\nu}}}\right) \right] ,\\
 \psi_{2}^{(\otimes)}(x)&= \left[\psi_{1}^{(\otimes)}(x)\right]^*
 \label{eq:jv} ,
\end{align}
where $\xi_{\tilde{\nu}}=\hbar/\sqrt{4m\tilde{\nu}}$ is the $\tilde{\nu}$-dependent length scale of the Josephson vortex.
The symbol $\otimes$ denotes the clockwise orientation of currents of the Josephson vortex (positive in component $\psi_{1}^{(\otimes)}$ and negative in $\psi_{2}^{(\otimes)}$), symbolically represented as $\uparrow \otimes \downarrow$. A degenerate solution is the \emph{(Josephson) anti-vortex} 
\begin{align}
\psi_i^{(\odot)} = \left[\psi_{i}^{(\otimes)}(x)\right]^* ,
\end{align}
for $i=1,2$, which has the opposite sense of rotation and is represented as
$\downarrow \odot \uparrow$. In the following we will use the term Josephson 
vortex for both vortex $\psi_{i}^{(\otimes)}$ and anti-vortex 
$\psi_{i}^{(\odot)}$ solutions, unless the orientation matters.
The solitary wave energy of the Josephson vortex (and the anti-vortex) evaluates to 
\begin{align}
E_s^\mathrm{JV}(\tilde{\nu}) 
%= \frac{8 \sqrt{\tilde{\nu}}(3 gn -4 \tilde{\nu})\hbar}{3g\sqrt{m}}  
= 2 \sqrt{\frac{\tilde{\nu}}{gn}}\left(3 -4 \frac{\tilde{\nu}}{gn}\right) E_\mathrm{DS} .
\end{align}

A linear stability analysis reveals that the dark soliton stack (for $M=2$) is
dynamically stable in the regime where the Josephson vortex solutions do not 
exist ($\tilde{\nu} > gn/4$) but dynamically unstable for $0<\tilde{\nu} < 
gn/4$, where they are subject to decay into stable Josephson vortices 
\cite{Kaurov2005,Qadir2012, Sophie2017}. In the absence of coherent coupling, at 
$\tilde{\nu}=0$, and at $\tilde{\nu} = gn/4$, dark solitons are marginally 
stable, i.e.\ the frequency of the unstable mode vanishes. The critical value of 
the linear coupling $\tilde{\nu} = gn/4$ (or, equivalently, 
$\tilde{\nu}/\tilde{\mu} = \frac{1}{3}$) is further the location of a pitch-fork 
bifurcation were the Josephson vortex and anti-vortex solutions become identical 
to the dark soliton stack solution. This can be seen from the energy, since 
$E_s^\mathrm{JV}(gn/4)=2 E_\mathrm{DS}$ and also directly from the wave function 
\eqref{eq:jv}.

Another pitch-fork bifurcation takes place at 
$\tilde{\nu}/\tilde{\mu} \approx 0.1413$ (or $\tilde{\nu}/gn \approx 0.1238$) 
where two additional stationary solitary wave solutions split off from each of 
the Josephson vortex and anti-vortex solutions \cite{Sophie2017}. These asymmetric 
solutions were referred to as maxima of the Josephson vortex dispersion relation 
in Ref.~\cite{Sophie2017}. Their essential character is revealed in the uncoupled limit ($\nu\to 0$) where a dark soliton remains in only one of the components results. 
At finite coupling $\nu>0$ one BEC component has a stronger density depression and steeper phase gradients while the other component displays a backflow current, while generally the currents are weaker than in the Josephson vortex solutions. 
We represent such  \emph{half-dark-soliton}  solutions schematically with $\node\downarrow$, where the arrow indicates the orientation of the (weak) backflow current. 
Specifically, the vortex solution $\uparrow\otimes\downarrow$ splits off two 
degenerate \emph{half-dark-soliton} solutions with signatures  $\node\downarrow$ and $\uparrow\node$. The anti-vortex solution $\downarrow\odot\uparrow$ analogously gives rise to half-dark-soliton solutions with signatures  
$\node\uparrow$ and $\downarrow \node$. The half-dark-soliton solutions may be 
thought of being obtained by moving the vortex position from between the two BEC 
components onto one of them. They coexist with the stationary Josephson-vortex 
solutions \eqref{eq:jv} for  $\tilde{\nu}/\tilde{\mu} \lessapprox 0.1413$, have 
higher energy, but also different canonical momentum and phase difference. The 
$\tilde{\nu}$-dependent energy $E_s^\mathrm{JV(M)}(\tilde{\nu})$ was computed in 
Ref.~\cite{Sophie2017}. In the region where they exist, the stationary 
half-dark-solitons and Josephson vortices are all dynamically stable due to 
their position on the \textit{yrast} dispersion relation \cite{Sophie2017}. 

In the remainder of this paper we consider stationary solitary wave solutions for $M>2$ BECs.

\subsection{Constructing solitary wave arrays from $M=2$ solutions}\label{sec:rep}

The known solutions for $M=2$ coupled condensates can be used to generate solutions for larger arrays with $M>2$ for certain symmetric configurations. In particular the dark soliton stack as written in Eq.\ \eqref{eq:dark} is a solution of the coupled GPEs \eqref{eq:gpe} for arbitrary $M$ and for both the planar and ring configuration as defined in Sec.\ \ref{sec:GPmodel}.

For an even number, $M=2r$, of condensates, two types of solutions of Eqs.\ \eqref{eq:gpe} can be constructed from known $M=2$ condensate solutions $\psi_1=A, \psi_2=B$ defined by Eqs.\ \eqref{eq:jv}:
\begin{enumerate}
\item \emph{Flipped repetition:} The third and fourth component repeat components one and two in reverse order. The pattern can be extended for any even number of condensates.
\begin{align}
(\psi_1,\psi_2,\psi_3,\ldots ,\psi_{2r}) = (A,B,B,A,A,B,\ldots)
%(\psi_1,\psi_2,\psi_3,\ldots ,\psi_{2r}) = (\psi_1^{(2)},\psi_2^{(2)},\psi_2^{(2)}, \psi_1^{(2)},\psi_1^{(2)},\ldots)
\end{align}
This solution is available for both the ring configuration and the planar configuration with $\nu= \tilde{\nu}$ and $\mu + \nu = \tilde{\mu}$. The reason why this works is that each condensate in the elementary $(A,B)$ unit is coupled to the nearest neighbor, which is identical. Therefore this linear coupling can be absorbed in adjusting the chemical potential $\tilde{\mu}$ of the two-component equation \eqref{eq:gpe2}.

\item \emph{Alternating repetition:} The third and fourth component repeat components one and two in the same order. The pattern can be extended for any even number of condensates.
\begin{align}
(\psi_1,\psi_2,\psi_3,\ldots ,\psi_{2r}) = (A,B,A,B,A,B,\ldots)
\end{align}
This solution is only available for the ring configuration with $2\nu = \tilde{\nu}$ and $\mu = \tilde{\mu}$. The reason why this works is that every condensate is coupled to two identical condensates, and thus the periodic $M$-condensate equation can be mapped onto the two-condensate GPE \eqref{eq:gpe2}. 
\end{enumerate}

The solitary wave energy is additive for repeated configurations and thus one
will obtain the $r$th multiple of the $M=2$ solitary wave energy. Let us 
specifically look at the possible solutions obtained for $M=4$.

\emph{Planar configuration of $M=4$ condensates.}
Flipped repetition of the Josephson vortex \eqref{eq:jv} produces the configuration  
\begin{align}
a = \;\uparrow \otimes \downarrow \; \downarrow \odot \uparrow
\end{align}
with the explicit solution 
\begin{align}
\nonumber
\psi^{(a)}_1(x) = \psi^{(a)}_4(x) &= \psi_{1}^{(\otimes)}(x), \\
\psi^{(a)}_2(x) = \psi^{(a)}_3(x) &= \left[\psi_{1}^{(\otimes)}(x)\right]^*,
\end{align}
with  $\nu= \tilde{\nu}$ and $\mu + \nu = \tilde{\mu}$. Note that flipping the
Josephson vortex in components 3 and 4 produces a Josephson anti-vortex. The 
energy of this configuration is
\begin{align}
E^{(a)}_s = 2 E_s^\mathrm{JV}(\nu) = 4 \sqrt{\frac{{\nu}}{gn}}
\left(3 -4 \frac{{\nu}}{gn}\right) E_\mathrm{DS} .
%\frac{16 \sqrt{{\nu}}(3 gn -4 {\nu})\hbar}{3g\sqrt{m}} .
\label{eq:a}
\end{align}
A degenerate solution with reversed currents 
$a^* = \; \downarrow \odot \uparrow\;\uparrow \otimes \downarrow$ is obtained by 
the complex conjugate of solution $\psi^{(a)}_i$.

The half-dark-soliton solutions of two-coupled condensate similarly
produce solutions for $M=4$ condensates by flipped repetition. Four solutions 
can be constructed:
\begin{align}
b_1 & = \node \downarrow\;\downarrow \node ,\\
b_2 & = \;\,\uparrow \node \node \uparrow, \\
b_3 & = \;\,\downarrow\node \node \downarrow ,\\
b_4 & = \node \uparrow \; \uparrow \node .
\end{align}
These solutions are degenerate with energy $E_s^{(b)} = 2 E_s^\mathrm{JV(M)}({\nu})$.

\emph{Ring configuration of $M=4$ condensates.}
In ring configuration, periodic boundary conditions apply and additional
coupling is possible between $\psi_1$ and $\psi_4$. The same solutions $a, a^*, 
b_i$ obtained for the planar configuration by flipped repetition are possible, 
since  $\psi_1 = \psi_4$ in these configurations. The solution $a^*$ is depicted in Fig.~\ref{fig:schem}.
From each of the known solutions, we can 
construct one new (degenerate) solution by cyclic permutation $\psi'_i = 
\psi_{i-1}$, e.g.
\begin{align}
a' =  \uparrow\;\uparrow \otimes \downarrow \; \downarrow \odot ,
\end{align}
where the last anti-clockwise vortex $\odot$ connects component $\psi_4$ with $\psi_1$. 

Additional solutions can be constructed from alternating repetition.
Applying this principle to the Josephson vortex solution  \eqref{eq:jv}, we obtain the solution 
\begin{align}
c = \;\,\uparrow \otimes \downarrow \odot \uparrow \otimes \downarrow \odot .
\end{align}
The explicit solution reads
\begin{align}
\nonumber
\psi^{(c)}_1(x) = \psi^{(c)}_3(x) &= \psi_{1}^{(\otimes)}(x), \\
\psi^{(c)}_2(x) = \psi^{(c)}_4(x) &= \left[\psi_{1}^{(\otimes)}(x)\right]^* ,
\end{align}
$2\nu = \tilde{\nu}$ and $\mu = \tilde{\mu}$. 
The energy is also given by twice the Josephson vortex energy, but now with a 
different interpretation of $\tilde{\nu}$ in terms of the coherent coupling 
strength of the $M=4$ BEC array. We obtain
\begin{align}
E^{(c)}_s =  2 E_s^\mathrm{JV}(2\nu) = 4 \sqrt{\frac{{2 \nu}}{gn}}\left(3 -8 \frac{{\nu}}{gn}\right) E_\mathrm{DS} .
%\frac{16 \sqrt{{2\nu}}(3 gn -8 {\nu})\hbar}{3g\sqrt{m}} .
\label{eq:c}
\end{align}
Applying alternating repetition to the Josephson anti-vortex solution yields a
degenerate configuration 
\begin{align}
 c^* =  \;\, \downarrow \odot \uparrow \otimes \downarrow \odot \uparrow \otimes,
\end{align}
which can alternatively be obtained by complex conjugation or cyclic permutation of $c$.

Alternating repetition of half-dark-soliton solutions yields another four
degenerate configurations with energy 
$E_s^{(d)} = 2 E_s^\mathrm{JV(M)}(2\nu)$:
\begin{align}
d_1 & = \node \downarrow \node \downarrow ,\\
d_2 & = \;\,\uparrow \node \uparrow \node , \\
d_3 & = \;\,\downarrow\node  \downarrow \node ,\\
d_4 & = \node \uparrow  \node \uparrow .
\end{align}

\subsection{Other solitary wave solutions}\label{sec:other}

The solitary waves constructed in Sec.~\ref{sec:rep} by flipped or alternating
repetition do not exhaust the  stationary solitary waves in the $M$-BEC array. 
In particular, the method described above does not allow us to construct 
solitary wave solutions in odd-$M$ arrays. But also for even $M$, the simplest 
non-trivial case being $M=4$, there are inaccessible solutions. One example is 
the single Josephson vortex solution
\begin{align} \label{eq:sv}
\uparrow \; \uparrow \otimes \downarrow \downarrow ,
\end{align}
which should exist as a stationary and stable solution in the planar
configuration by analogy to the solitonic vortex solution known to exist in 
finite-width 2D and 3D waveguide geometries for BECs 
\cite{Brand2002,Mateo2015a,Toikka2016}. Other, less symmetric solitary waves 
consist as well and can be found by numerical procedures.

From a topological perspective, configurations like the solitonic
vortex analog \eqref{eq:sv} with an unbalanced number of Josephson vortices 
cannot exist as finite-$E_s$ solitary waves in the ring configuration, where 
vortex lines that enter the ring also have to leave it. In the following section 
we will study the stability of solitary waves in the BEC array, which will also 
allow us to identify bifurcation points.

\section{Linear stability of the dark soliton stack in ring configuration} \label{sec:Bog}

In order to facilitate analytic results, we consider the ring configuration in the remainder of the paper.
The dark soliton stack \eqref{eq:dark} is a stationary solution of the coupled
GPEs \eqref{eq:tgpe}. In order to learn about its dynamical stability 
properties  and bifurcation points we follow the well-know procedure of Refs.~\cite{Kuznetsov1988,Muryshev1999,Mateo2015a,Mateo2014} to linearize the 
time-dependent GPE \eqref{eq:tgpe} around the dark soliton solution  
\eqref{eq:dark} with the ansatz
$\Psi_j(x,t) = [\psi_j(x)+u_j(x) e^{-i\omega t} +v^*_j(x) e^{i\omega t}]e^{-i\mu t/\hbar}$. 
We thus obtain the set of coupled Bogoliubov  equations for the linear
excitation modes with frequency $\omega$ and amplitudes $u_{j}(x),v_{j}(x)$ 
describing the small-amplitude excitations of the BEC $\Psi_j$
\begin{align}
 H_j \, u_j+  \, g  \, n \, \tanh^2\left(\frac{x}{\xi}\right) 
v_j 
-\nu \left(u_{j-1}+u_{j+1}\right) & = \hbar \omega \, u_j \nonumber\\
 -H_j \, v_j -  \, g  \, n \, \tanh^2\left(\frac{x}{\xi}\right) u_j 
+\nu  \left(v_{j-1}+v_{j+1}\right) &= \hbar \omega \, v_j \,,
\label{eq:bog0}
\end{align}
where $H_j = -(\hbar^2/2m)\partial_{xx}  +  2g n  \tanh^2(x/\xi)-\mu$.

Since the dark soliton solutions \eqref{eq:dark} are real-valued wave functions,
it is convenient to re-arrange Eqs.\ (\ref{eq:bog0}) for  the symmetric and 
antisymmetric combinations $f_{j\pm}=u_{j}(x) \pm v_{j}(x)$. 
In order to account for the $j$ variation across the BEC 
array, we expand these modes in a plane wave basis along the 
$j$-direction (i.e.\ along the vertical direction in 
Fig.\ \ref{fig:schem}), that is
\begin{equation}
 f_{j\pm}(x)=\sum_{k} f_{\pm}^{(k)}(x) e^{i k \xi_y j} \,,
 \label{eq:fourier}
\end{equation}
where we have defined $\xi_y=\sqrt{\hbar^2/2m\nu}$ as a length scale associated with the 
linear coupling $\nu$.
We thus obtain
\begin{align}
 \left[\frac{-\hbar^2}{2m}\partial_{xx}-  \, g \,n \, 
\mbox{sech}^2\left(\frac{x}{\xi}\right) + \lambda_k \right] f_{-}^{(k)}&= \hbar \omega
f_{+}^{(k)}
\label{eq:bog1-}
\\
  \left[ \frac{-\hbar^2}{2m}\partial_{xx} -  3\, g\, n \, 
\mbox{sech}^2\left(\frac{x}{\xi}\right) + \lambda_k  \right] f_{+}^{(k)}&=\hbar\omega
f_{-}^{(k)} \,,
\label{eq:bog1+}
\end{align}
where 
\begin{equation}
 \lambda_k=2\nu \left[1- \cos(k \xi_y)\right] \,,
  \label{eq:dispersion}
\end{equation}
provides the dispersion relation along the $j$ ``coordinate'', which can be seen
as a synthetic second dimension (where we may consider $y = 
j \xi_y$ a synthetic coordinate with the dimension of length). 
Because of the periodic arrangement of BECs, the wave number $k$ can only take 
$M$ values within the first Brillouin zone,
\begin{equation}
 \frac{M\xi_y}{2\pi}\,k = 0, \pm 1, \pm 2, \dots, \frac{M}{2} \, ,
 \label{eq:ks}
\end{equation}
and  the transverse energy Eq.\ (\ref{eq:dispersion}) represents a discrete 
spectrum. 

The coupled Eqs.\  (\ref{eq:bog1-})--(\ref{eq:bog1+}) [or equivalently
Eqs.\ \eqref{eq:bog0}] present a linear non-hermitian eigenvalue problem and 
thus permit, in principle, complex  eigenvalues $\omega$. It is easy to see 
[e.g.\ by eliminating $f_+^{(k)}$ from 
Eqs.\  (\ref{eq:bog1-})--(\ref{eq:bog1+})] that $\omega^2$ is real-valued.
Thus the possible solutions for $\omega$ are either real, signifying stable 
normal modes, or purely imaginary, signifying dynamically unstable linear modes. 
At the transition between stable and unstable regions in parameter space, 
necessarily at least one normal mode frequency has to pass through zero. For 
this reason we solve Eqs.\  (\ref{eq:bog1-})--(\ref{eq:bog1+}) for $\omega=0$ 
in order to obtain the stability boundaries.

In the absence of transverse excitations, i.e.\ for $k=0$ and then 
$\lambda_k=0$, the Bogoliubov equations (\ref{eq:bog1-})--(\ref{eq:bog1+}) have 
zero frequency modes (with $\omega=0$)  that arise from the continuous 
symmetries of the system, the Goldstone modes. The Eqs.\  
(\ref{eq:bog1-})--(\ref{eq:bog1+}) furthermore decouple for $\omega=0$ and can 
be solved independently.
Specifically, Eq.\  (\ref{eq:bog1-}) is solved by 
$f_{-}^{(0)}= \exp(i\theta)\sqrt{n}\,\tanh({x}/{\xi}) $, reflecting the symmetry 
of the stationary solution 
(\ref{eq:dark})  against the selection of a global phase $\theta$. Equation\  (\ref{eq:bog1+}) admits the Goldstone mode solution 
$f_{+}^{(0)}=\xi \,\partial_x 
\sqrt{n}\,\tanh({x}/{\xi})=\sqrt{n}\,\mbox{sech}^2({x}/{\xi}) $,
which is localized at the density notch and corresponds to the invariance of 
Eq.\ (\ref{eq:dark}) against translations along the axial coordinate.

\begin{figure}[t!]
\includegraphics[width=8.5cm]{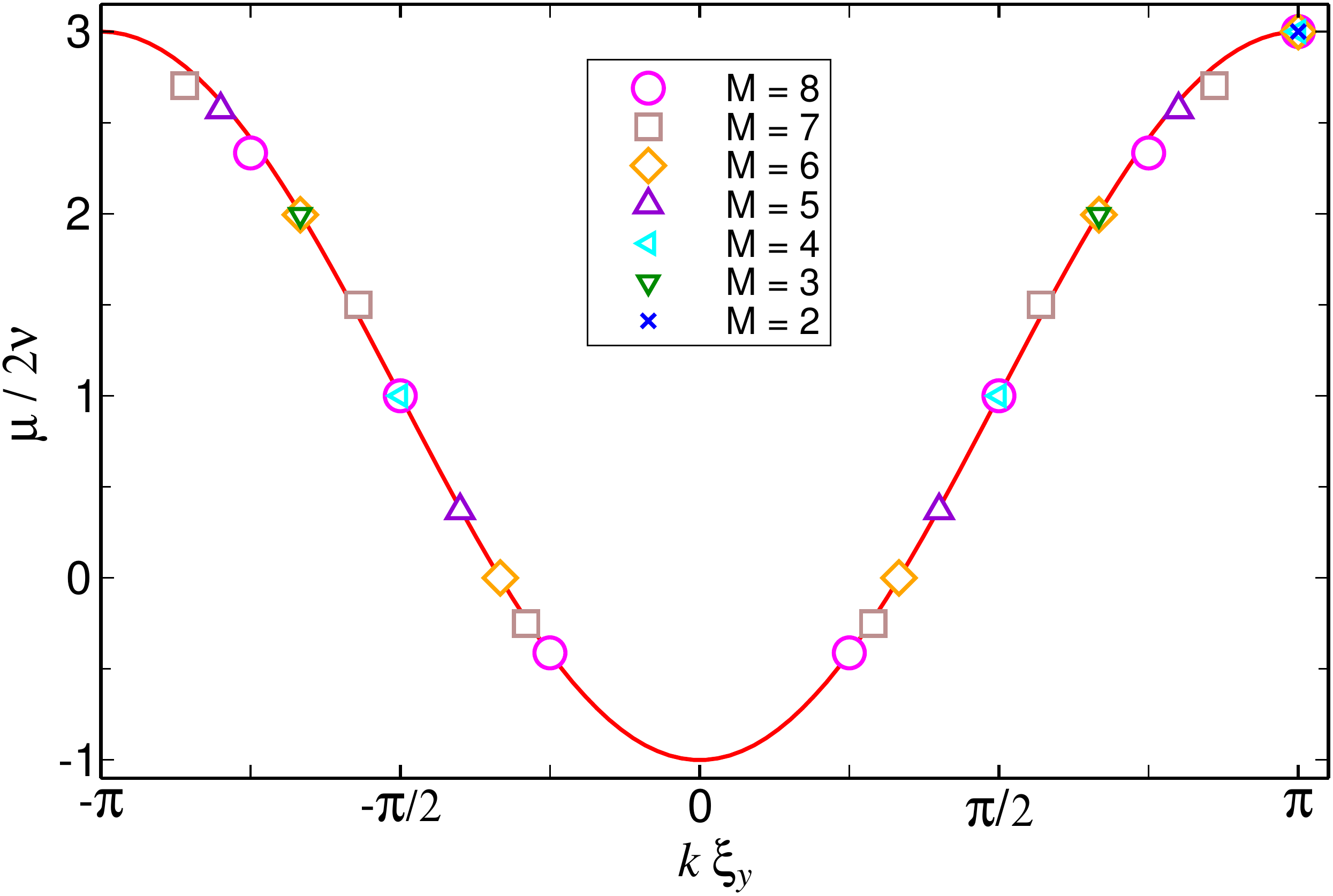}
\caption{Bifurcation points from the stationary dark solitons in systems with 
$M=2,\dots,8$ linearly coupled BECs in a ring configuration. Transverse modes 
with discrete momentum $k$ are excited along the effective dimension $y$ 
introduced by the coherent coupling $\nu$ (see text). The solid curve 
represents the analytical expression (\ref{eq:bif}) and the open symbols 
correspond to the numerical solutions of Bogoliubov equations  (\ref{eq:bog0}). 
}
\label{fig:bifurcation}
\end{figure}  

In addition to the Goldstone modes, which arise from symmetries and thus are
always present, the Bogoliubov equations (\ref{eq:bog1-})--(\ref{eq:bog1+}) admit 
zero frequency ($\omega=0$) solutions under certain conditions, e.g.\ when 
normal modes become unstable, which is exactly where new nonlinear solutions 
bifurcate. We can find such solutions in the presence of transverse 
excitations, i.e.\ for $k \neq 0$. In particular, note that Eq.\ 
(\ref{eq:bog1-}) with $\omega=0$ is a Schr\"{o}dinger equation for $f_{-}$ with 
a potential well defined by $V(x)=- \, g \,n \, \mbox{sech}^2({x}/{\xi})$. The 
nodeless and localized (bound) ground state solution is known from the 
literature  to take the form
\begin{equation}
  f_{-}^{(k)}(x)=\sqrt{n}\, \mbox{sech}\left(\frac{x}{\xi}\right) \,,
 \label{eq:fbif}
\end{equation}
with eigenvalue $-g \,n/2$ \cite{Rosen1932}. This becomes a valid zero-frequency
solution of the Bogoliubov equations, if the eigenvalue matches $-\lambda_k$, 
i.e.\ $\lambda_k=g \,n/2=(\mu+2\nu)/2$.
Rewriting this equation as a condition relating the chemical potential and the
linear coupling yields
\begin{equation}
 \mu_{b}=2\nu \left[1- 2\cos({k}\xi_y)\right]  \,.
 \label{eq:bif}
\end{equation}
This expression, which can be matched at the $M-1$ discrete values 
${k}\xi_y=(\pm 1, \pm 2,...,{M}/{2})\times 2\pi/M$, indicates 
the existence of corresponding bifurcation points in the family of dark soliton 
states as a function of $\mu$ (or equivalently, as a function of $\nu$). 
Figure \ref{fig:bifurcation} shows a comparison of the analytical values 
given by Eq.\ (\ref{eq:bif}) with the results  obtained by  
numerically solving the Bogoliubov equations (\ref{eq:bog0}) for $M=2,\dots,8$. 
The tiny disagreement in the numerical values comes from a discretization
error in the numerical data.

\begin{figure}[t!]
\includegraphics[width=8.5cm]{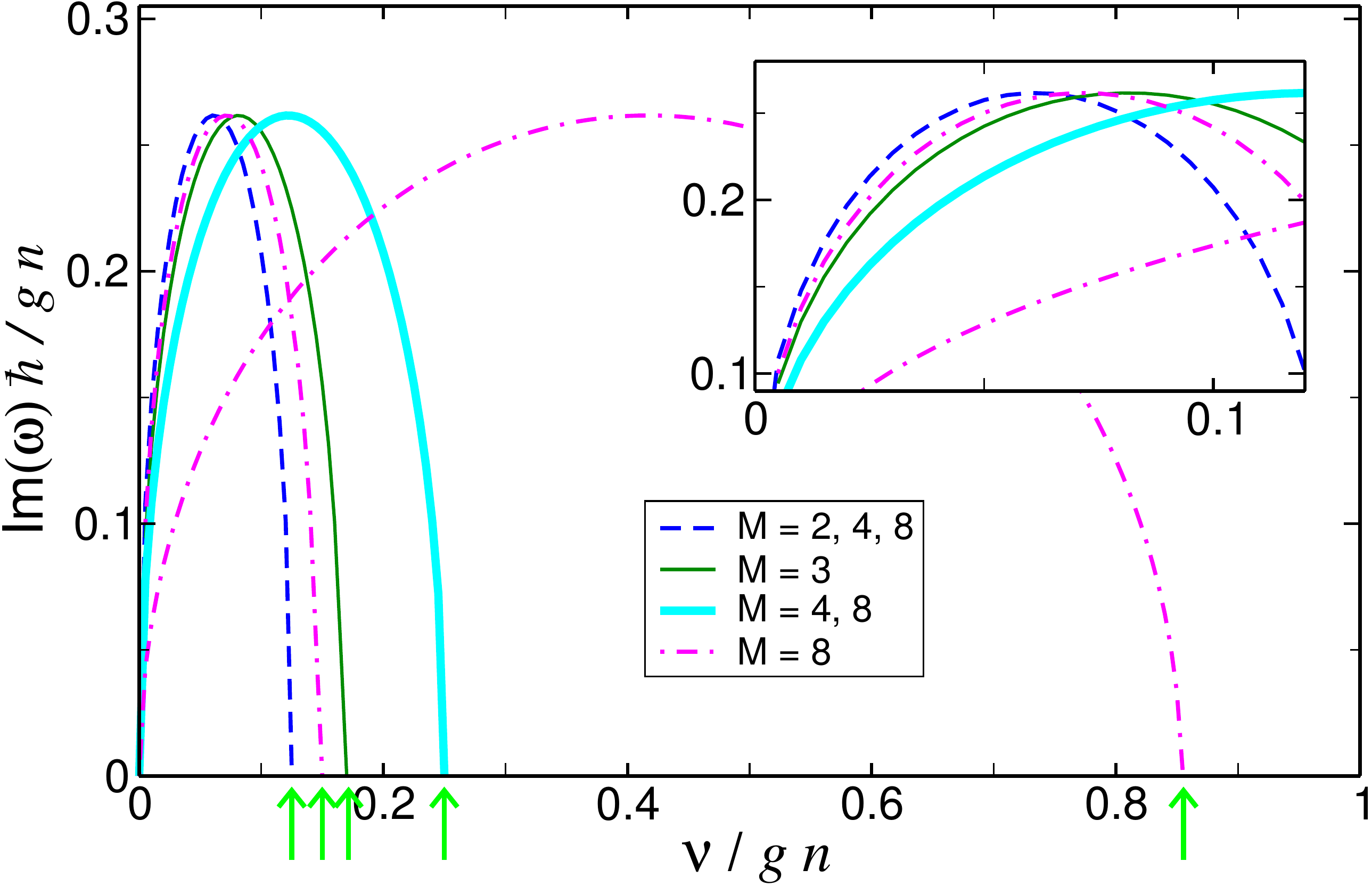}
\caption{Unstable excitation frequencies from the numerical solution of the 
Bogoliubov equations (\ref{eq:bog0}) for dark soliton states in stacks of 
$M=2,3,4,8$ condensates in a ring configuration and varying coupling 
energy $\nu$ [keeping  the background density $n=(\mu+2\nu)/g$ constant]. The 
arrows under 
the horizontal axis indicate the bifurcation points (represented in Fig.\ 
\ref{fig:bifurcation}), predicted by the analytical expressions (\ref{eq:bif}) 
and (\ref{eq:nubif}).}
\label{fig:unst_freq}
\end{figure}

Solving the full Bogoliubov equations  (\ref{eq:bog0}) numerically, reveals that
the zero-frequency normal modes of Eq.\ (\ref{eq:bif}) acquire imaginary 
frequency when $\mu$ is increased (or $\nu$ is decreased, 
respectively), signifying a dynamical instability of the dark soliton array. 
Figure \ref{fig:unst_freq} collects our numerical data for the unstable 
frequencies of dark solitons in different stacks ($M=2,3,4,8$)  with the same 
background density $n=(\mu+2\nu)/g$. The vertical arrows under the 
horizontal axis point to the predicted bifurcations given by Eq.\ 
(\ref{eq:bif}) and shown in Fig.\ \ref{fig:bifurcation}.
Now they are presented as a function of the linear coupling, after re-arranging 
Eq.\ (\ref{eq:bif}):
\begin{equation}
 \nu_{b}=  \frac{gn}{4\left[1-\cos({k}\xi_y)\right]}  \,.
 \label{eq:nubif}
\end{equation}
For arrays with an even number of condensates 
$M=2,4,8$, there are degenerate frequencies 
(and then also bifurcations) due to 
symmetries. Specifically, the same unstable modes present for $M=2$ reappear 
for $M=4$ where additional modes are present (and correspondingly for $M=8$).
Due to the limited range of $k$ values per Eq.\ \eqref{eq:ks}, the relation
between $\nu_{b}$ and $k$ is monotonic. Specifically, the unstable modes with 
larger $k$ values are associated with a smaller critical coupling   $\nu_b$. 
Since the instability regions overlap, the last critical coupling $\nu_b$ that 
marks the region between stability and instability for the dark solitons occurs 
for the smallest modulus of $|k|=2\pi/M\xi_y$. As a consequence, the dark soliton stack with overlapping dark solitons as per Eq.\ \eqref{eq:dark} is stable if
\begin{align}\label{eq:stabDS}
\frac{\nu}{gn} > \frac{1}{4\left[1-\cos(2\pi/M)\right]} .
\end{align}
As we discuss in the next 
paragraph, this last instability threshold also marks the bifurcation of a 
single pair of Josephson vortex and Josephson antivortex from the dark soliton solution.

\section{Bifurcations of Josephson vortices} \label{sec:JV}

\begin{figure}[t!]
\includegraphics[width=\linewidth]{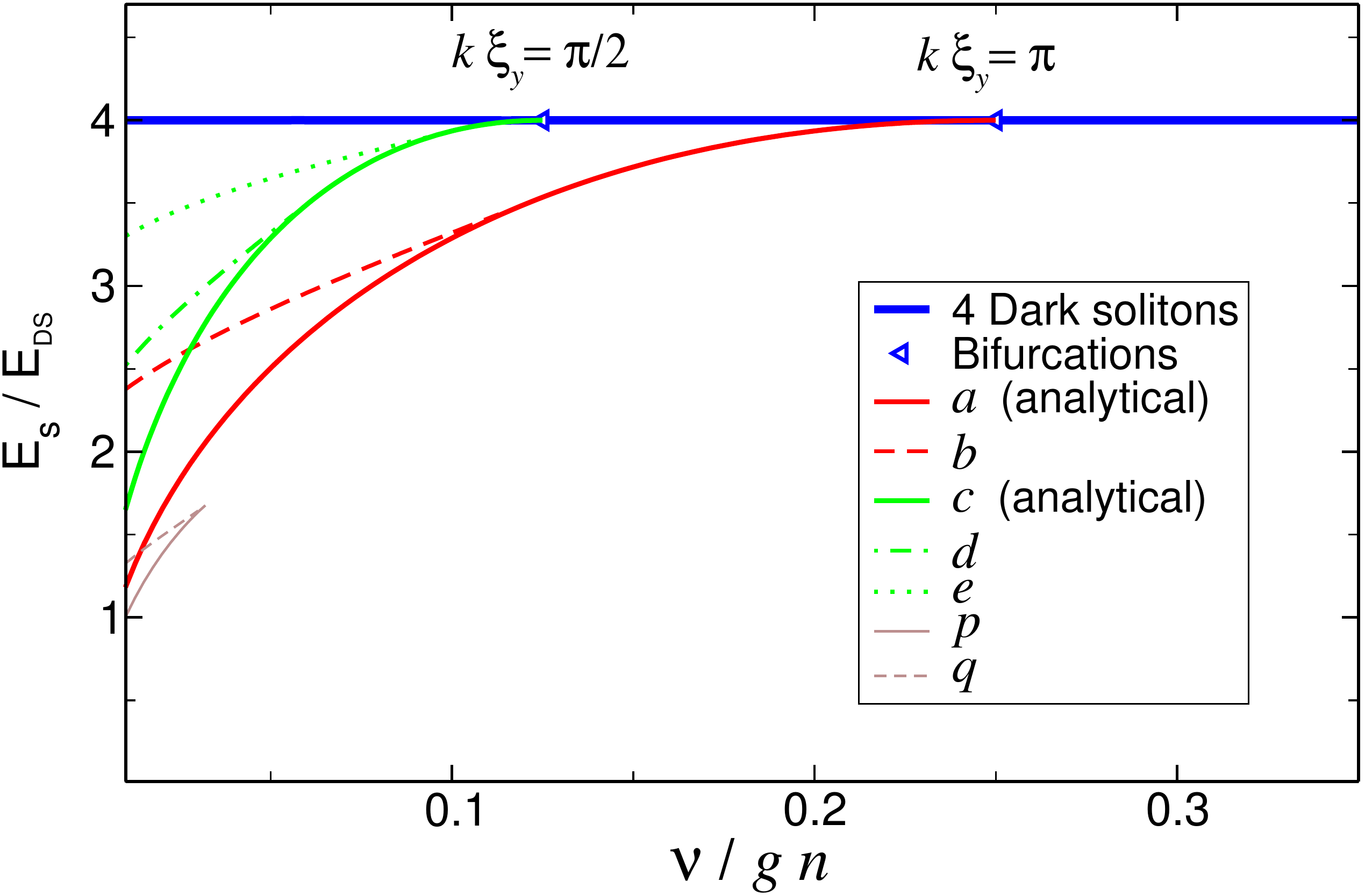}\\ \vspace{-0.7cm}
\includegraphics[width=\linewidth]{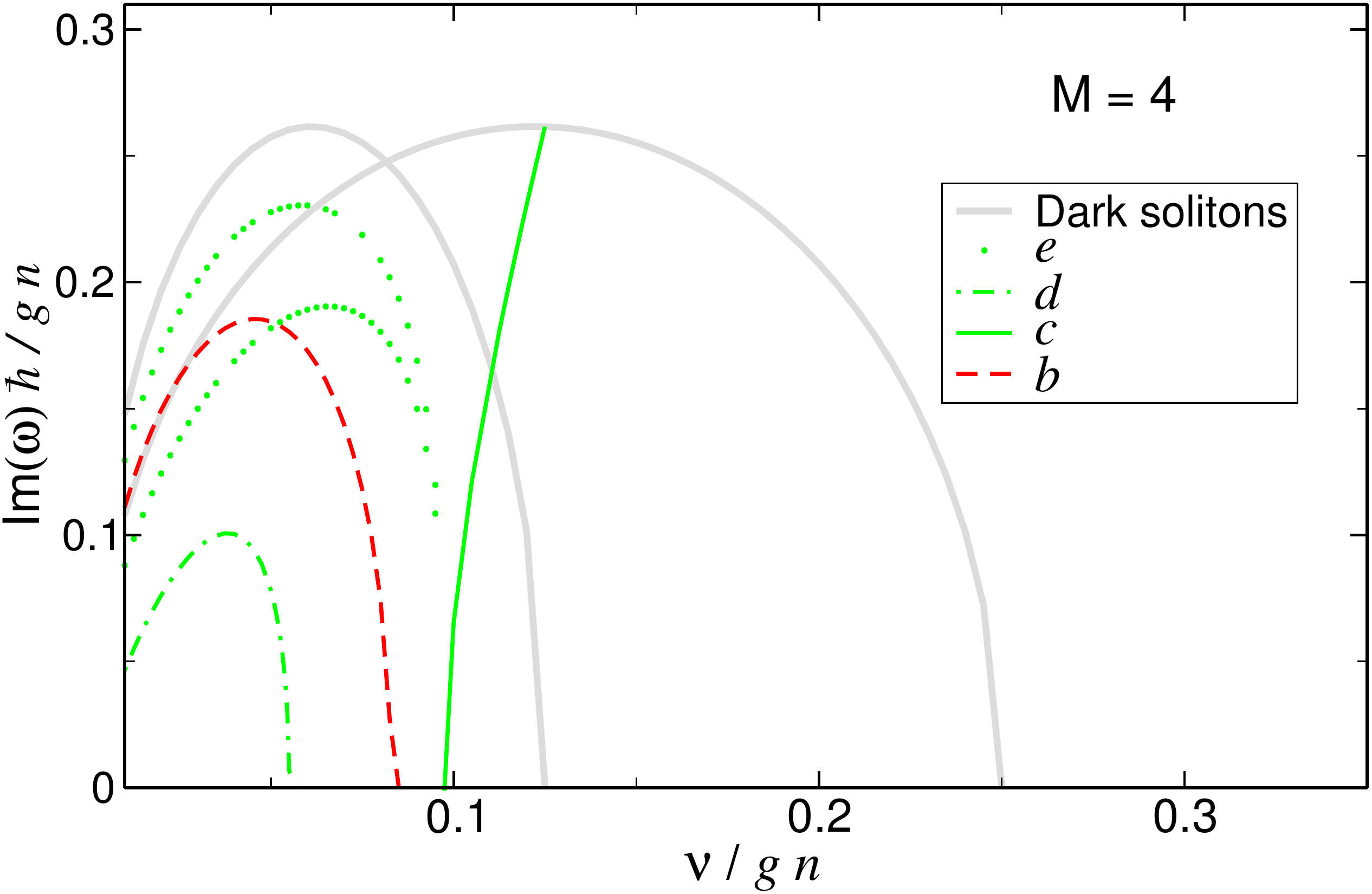}
\caption{Stationary solitary wave solutions in an array of $M=4$ BECs in  ring 
configuration. 
Top panel: excitation energy vs.\ the linear coupling strength for solitary 
waves on the constant background density $n=(\mu+2\nu)/g$.
The dark soliton stack of Eq.~\eqref{eq:dark} has constant energy with
Josephson-vortex-type solutions bifurcating at the parameter values indicated 
with open triangles. The corresponding wave-number value from 
Eq.~\eqref{eq:nubif} (and Figs.~\ref{fig:bifurcation} and \ref{fig:unst_freq}) 
is indicated for each bifurcation point. See Figs.~\ref{fig:JVs} and 
\ref{fig:JV1} for representative states in each branch.
Bottom panel: unstable frequencies of the Bogoliubov spectrum of the solitary 
waves in the top panel. The configurations $a$, $p$ and $q$ are always stable, $d$, and $e$ always unstable, while configuration $b$, $c$, and the dark soliton stack change stability in different regimes of $\nu/gn$.   
}
\label{fig:4comp_JVtraj}
\end{figure}

\begin{figure}[t!]
\includegraphics[width=\linewidth]{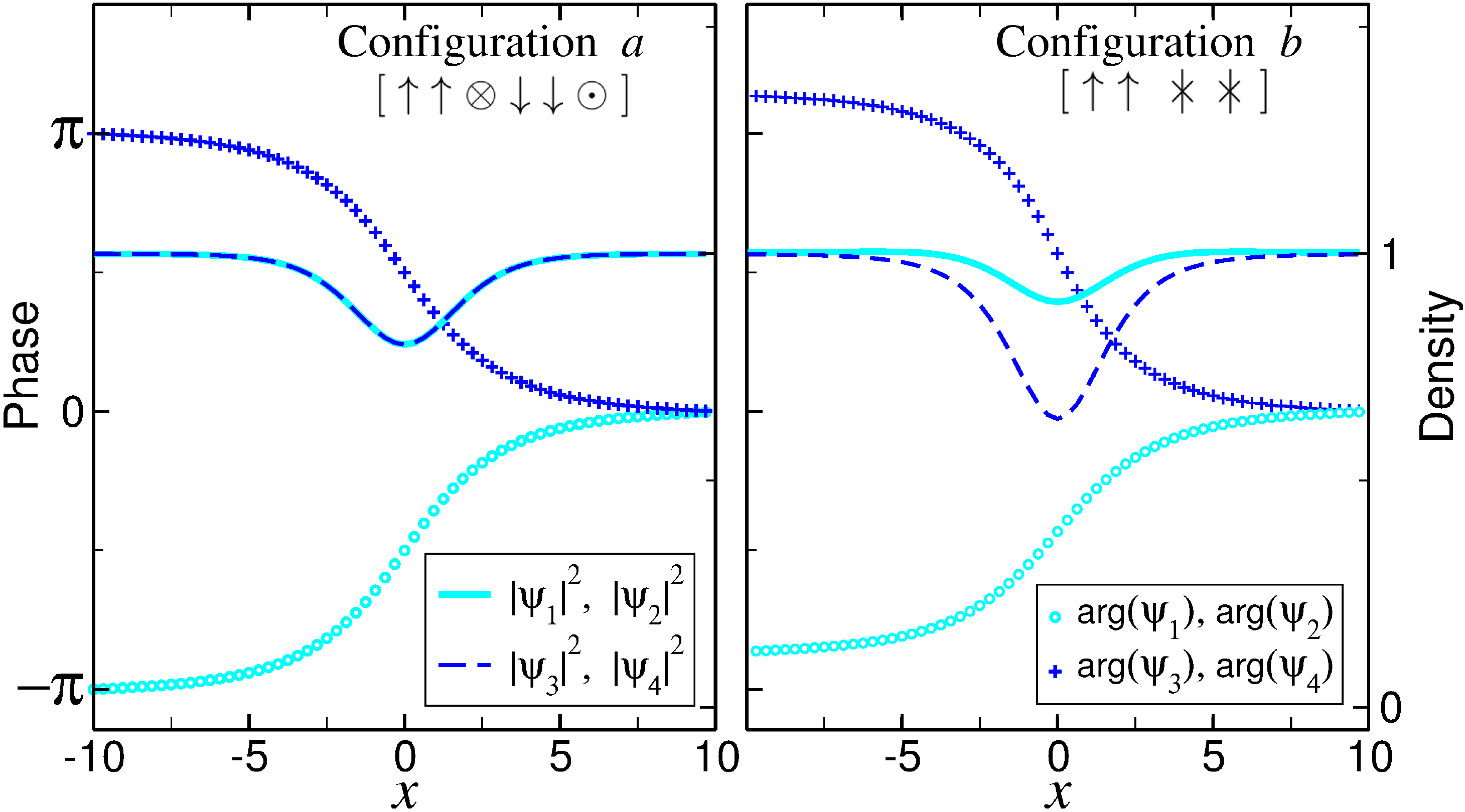}\\ \vspace{-0.5cm}
\includegraphics[width=\linewidth]{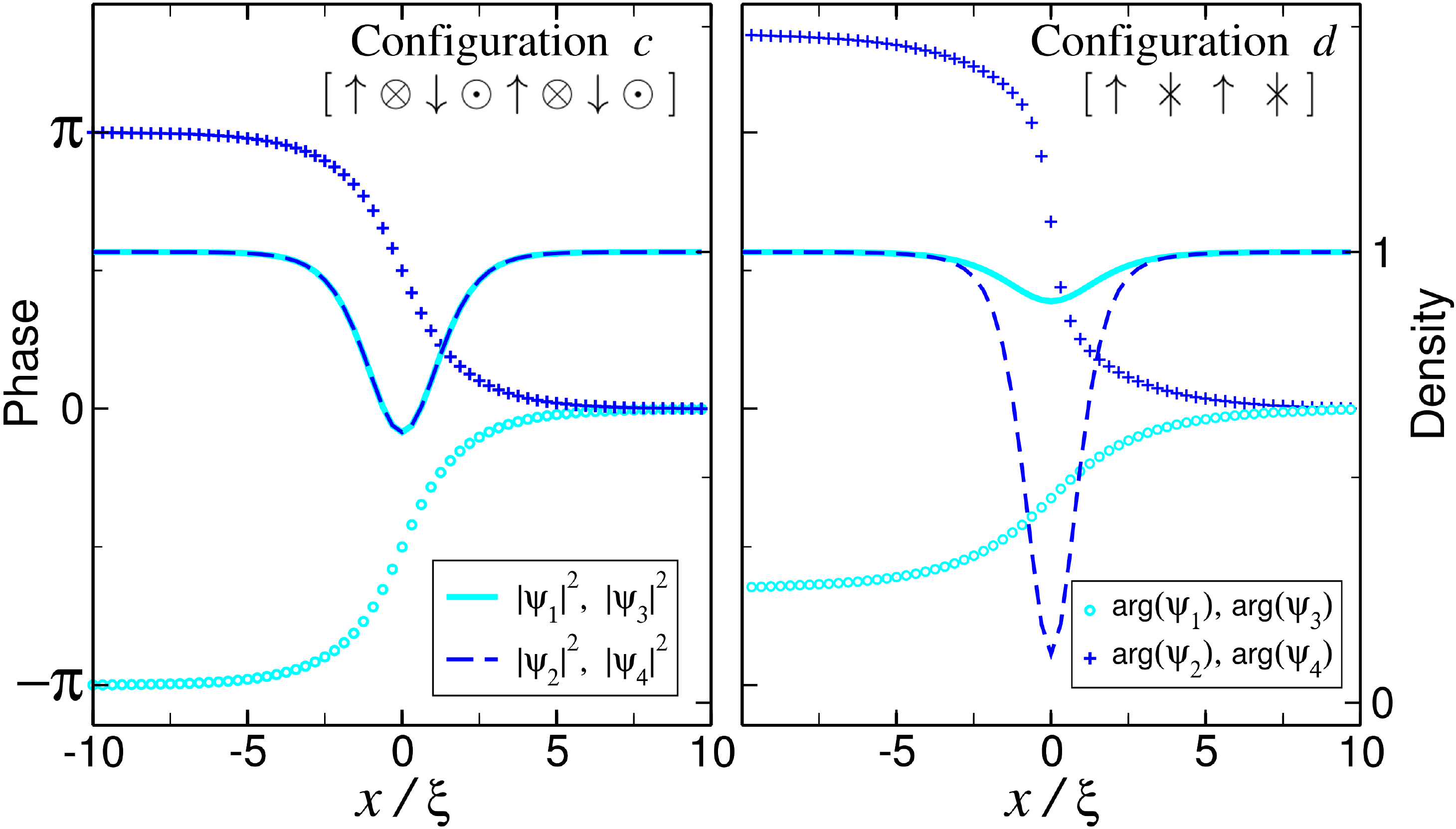}\\ \vspace{-0.6cm}
\includegraphics[width=\linewidth]{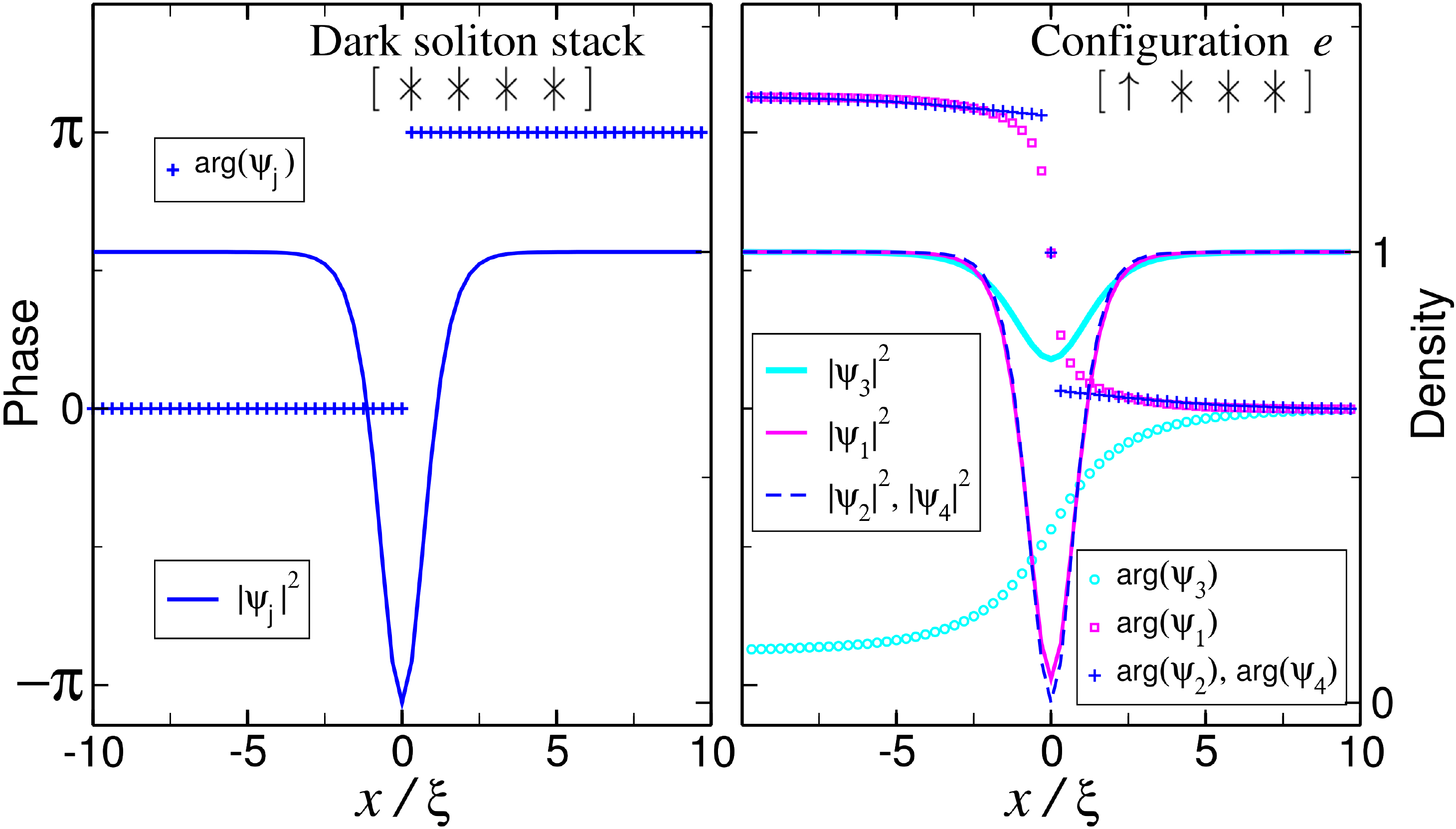}
\caption{Density and phase profiles of stationary solitary waves in a 4-BEC 
array in ring configuration with  $\nu=0.05 \,gn$: (symmetric) 
Josephson vortex solutions  $a$ and 
$c$, dark solitons, and fractional-dark-soliton
configurations $b$, $d$ and $e$. Legends are shared between panels in the top and middle rows.
Of the shown configurations only  $a$ and $c$ are dynamically stable (see 
text). Density is reported in units of $n$.}
\label{fig:JVs}
\end{figure}

The critical coupling values $\nu_b$, indicated by arrows in Fig.\
\ref{fig:unst_freq}, mark the bifurcations of new families 
of solitary waves from the dark soliton stack \eqref{eq:dark} in the ring configuration. As described in 
Sec.\ \ref{sec:rep}, these families are made of stationary Josephson vortices 
that break the 
time reversal symmetry of the GPE \eqref{eq:tgpe} and present lower 
excitation energies than the original dark solitons. Figure 
\ref{fig:4comp_JVtraj} shows the energy of the two emerging branches (continuous 
lines) of solitary waves bifurcating from the $M=4$ dark-soliton stack (thick 
horizontal line). 
The bifurcations take place at values of the coupling parameter $\nu_b=gn/4$ and $\nu_b=gn/8$, 
as obtained from Eq.\ \eqref{eq:nubif} at the momentum values 
$k\xi_y=\pm\pi$ and $k\xi_y=\pm\pi/2$. The states in these branches consist of, 
respectively, Josephson vortices in the configuration $a =[ \;\uparrow 
\otimes \downarrow \; \downarrow \odot \uparrow]$,  with energy given by Eq.\ 
\eqref{eq:a}, and the configuration $c = [\;\uparrow \otimes \downarrow 
\odot \uparrow \otimes \downarrow\odot] $, with higher energy than $a$, given by 
Eq.\ \eqref{eq:c}.
As these solutions explicitly break the symmetry under reversal of current and with respect to cyclic permutation of the array (translation along the periodic array), such operations generate degenerate solutions, as previously discussed. This yields a total of four degenerate solutions of type $a$ and two degenerate solutions of type $c$.
Representative states are depicted in Fig.\ \ref{fig:JVs}. In both cases the 
density profiles are equal for all the components. As the value of the coupling parameter $\nu$ is increased, the density dips in the condensates deepen within each family. The same is also the case for the other states 
containing Josephson vortices.

For decreasing values of the coupling, the density-symmetric families of 
Josephson vortices $a$ and $c$ experience in turn additional bifurcations 
(discontinuous 
lines in Fig. \ref{fig:4comp_JVtraj}) that give rise to states with non equal 
density profiles along the stack. In particular, the half soliton states $b=[ 
\node \downarrow\;\downarrow \node ]$ (dashed line) bifurcate from family $a$, 
whereas the half solitons $d=[ \node \downarrow\node\downarrow ]$ (dash-dotted 
line) bifurcate from family $c$.

Besides, there are other stationary solutions in the 4-BEC stack that were not 
accounted for in Sec. \ref{sec:rep}. First, there is an additional bifurcation 
of  
family $c$, taking place at higher energy than branch $d$, which 
correspond to configurations of the type $e=[ \node \node \node \downarrow]$ 
(dotted line in Fig. \ref{fig:4comp_JVtraj}). When the linear coupling vanishes 
the energy of this branch equals 3$E_{DS}$. 
Additionally, we have also found two branches of Josephson vortex type solutions 
that are disconnected from the families $a$ to $e$ presented before. They 
correspond to configurations  of the type $p=[ \uparrow\;\uparrow\;\uparrow 
\otimes\downarrow\odot]$ (thin continuous line) and 
$q=[\uparrow\;\uparrow\;\uparrow \node]$ (thin dashed line), whose 
energies tend to zero and $E_{DS}$, respectively, 
when $\nu\rightarrow 0$. They merge at the maximum energy value, 
corresponding also to the maximum coupling value in both families. 
 
\begin{figure}[t!]
\includegraphics[width=\linewidth]{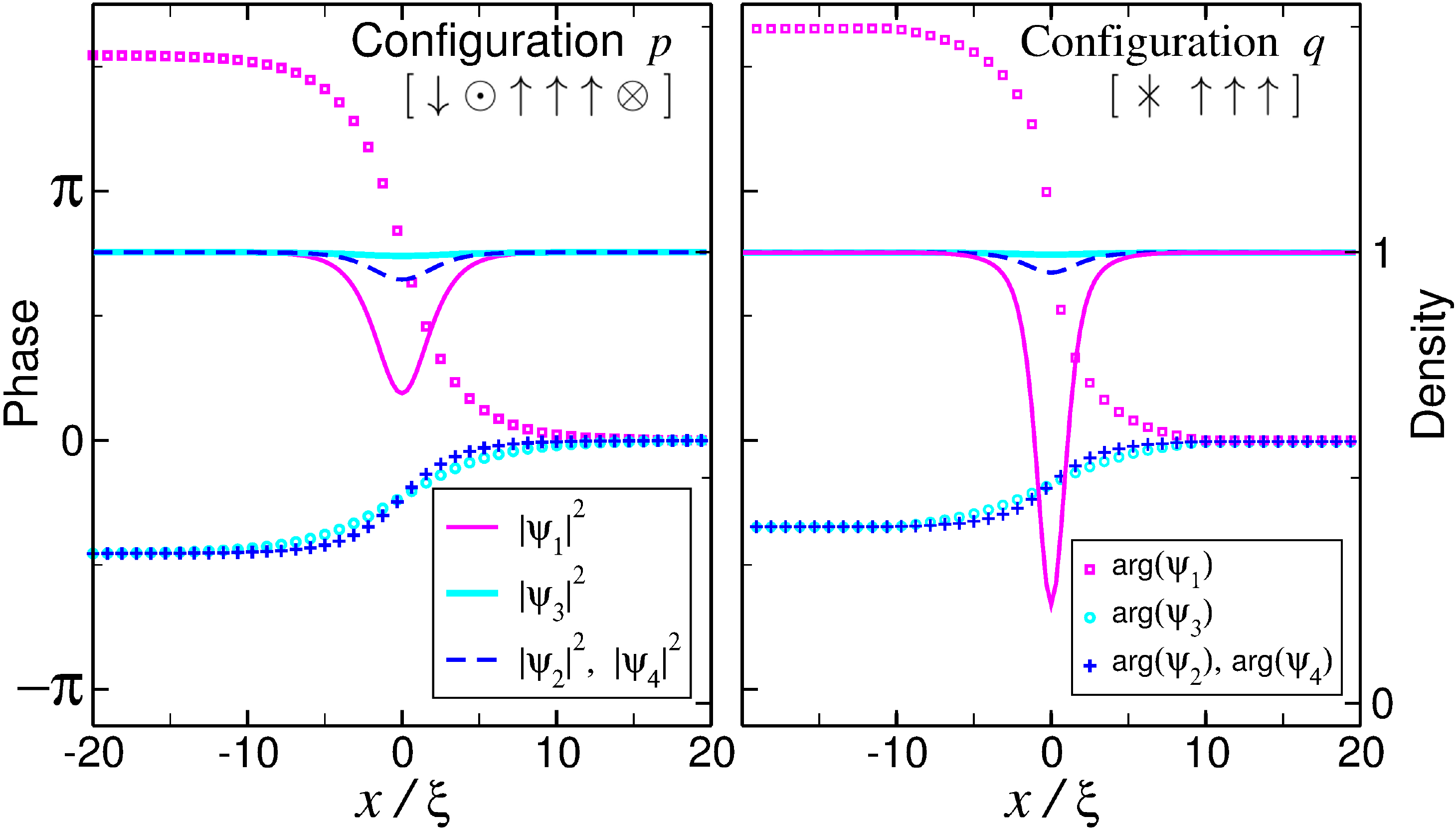}
\caption{Density and phase profiles of solitonic states $p$ 
and $q$ in an array of 4-BEC in ring configuration with  
$\nu=0.025 \,gn$. Legends are shared by both panels. These states are 
dynamically stable. Density is reported in units of $n$.}
\label{fig:JV1}
\end{figure}

In Fig.\ \ref{fig:JVs} we plot the density and the phase profiles of 
the solitonic states $a$-$e$ in the $M=4$ BEC array for the same value of the 
density and linear coupling $\nu=0.05\,gn$. For comparison, the dark soliton 
stack is also depicted in the bottom left panel. The profiles of configurations 
$p$ and $q$ are depicted in Fig.\ \ref{fig:JV1} for linear coupling 
$\nu=0.025\,gn$. As can be seen, these latter states, different from the other 
configurations, shape a density notch which does not extend to the whole array. 
In configuration $p$, a vortex dipole, made of a Josephson vortex and a Josephson antivortex, 
localizes around a single strand. This solution is continuously connected to configuration $q$, which represents a solitary wave that is largely concentrated on a single strand. These solutions $p$ and $q$ represent solitary waves that are not only localized in $x$ direction but along the transverse direction (i.e.\ along the array) and thus realize lump-type solitons. They very closely resemble the Jones-Roberts vortex-dipole and rarefaction-type solitary waves known in continuous two-dimensional BECs  \cite{Jones1982}, with the difference that the latter always move with finite velocity whereas the solutions represented in Figs.\ \ref{fig:4comp_JVtraj} -- \ref{fig:JV1} are stationary.

\subsection{Stability of bifurcating solitary waves}

Regarding the stability, we have numerically solved the 
Bogoliubov equations for the linear excitations of the Josephson vortex 
solutions considered in Fig.\ \ref{fig:4comp_JVtraj}. The frequencies of 
unstable excitations are plotted in the bottom panel of this figure. Among the 
states connecting with the dark soliton stack 
(through direct or secondary bifurcations)  only the family with configuration 
$a$, the flipped configurations of Josephson vortices, contains dynamically 
stable states for all the values of the coupling parameter. The situation is more 
complicated for the other branches. The family of states with 
configuration $c$ (alternating Josephson vortices) becomes only unstable  
from the point of the secondary bifurcation of branch $e$ up to its 
junction with the dark soliton stack. The family $b$ is also stable in a small 
range of couplings just before merging into its parent branch. Families $d$ and 
$e$ are always unstable. 
Interestingly, the two families $p$ and $q$, disconnected from the dark soliton 
stack, contain dynamically stable states. 

\subsection{Real time evolution}

\begin{figure}[t!]
\includegraphics[width=\linewidth]{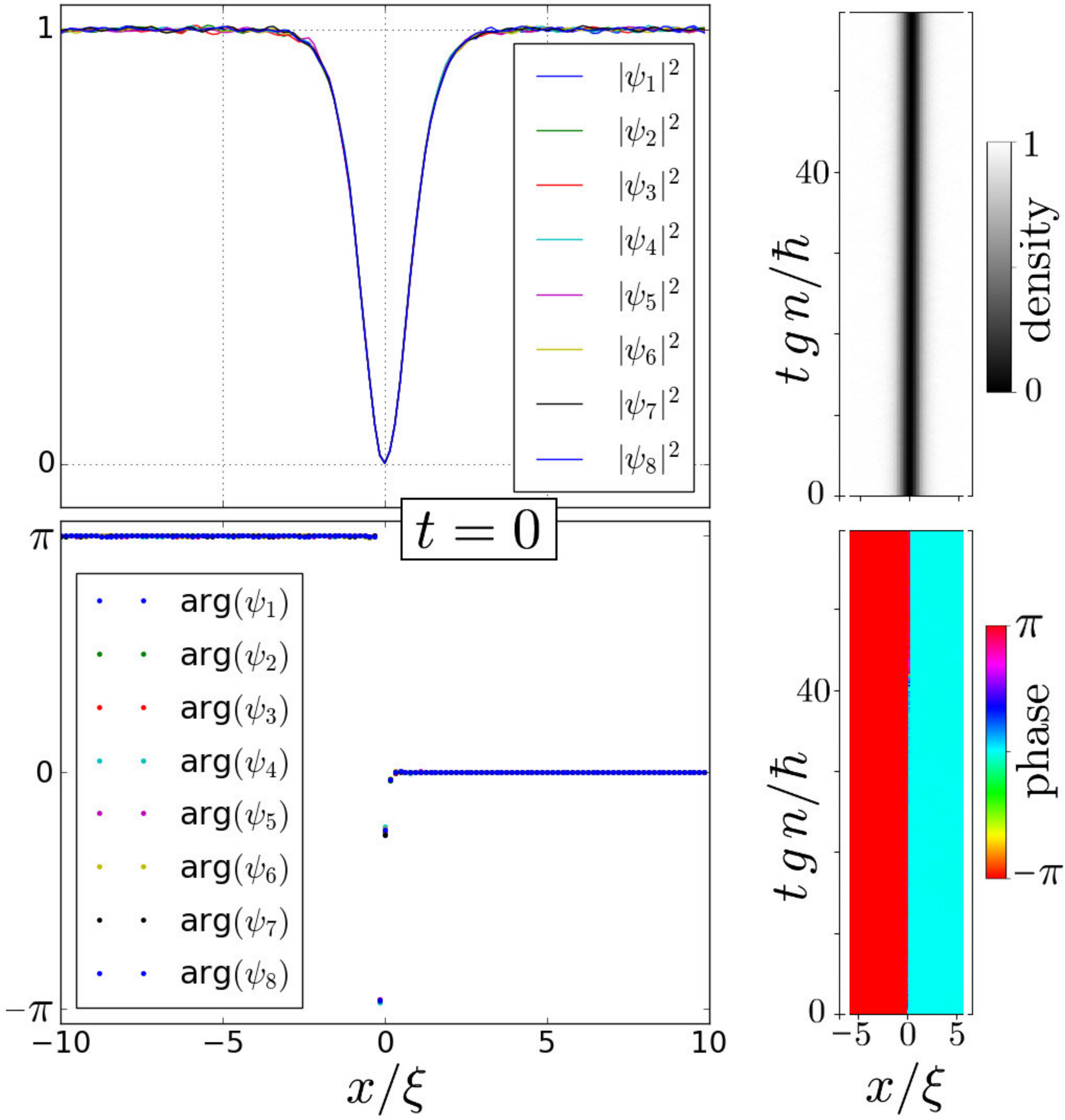}
\caption{Time evolution of a dark soliton stack with overlapping dark solitons in an array of $M=8$ BECs in ring configuration. The linear coupling parameter $\nu/gn =0.9$ is in the stable regime.
Left panels: Initial density and phase profiles after a 
$1\%$-amplitude white-noise perturbation added to the wave function of Eq.~\eqref{eq:dark}.  
Right panels: Density (top) and phase (bottom) for one component (all 
components follow almost identical evolutions) are shown 
around the initial soliton position (at $x=0$) as a function of time (vertical 
axis).  
Density is reported in units of $n$.
}
\label{fig:8comp_stable}
\end{figure}

\begin{figure}[t!]
\includegraphics[width=\linewidth]{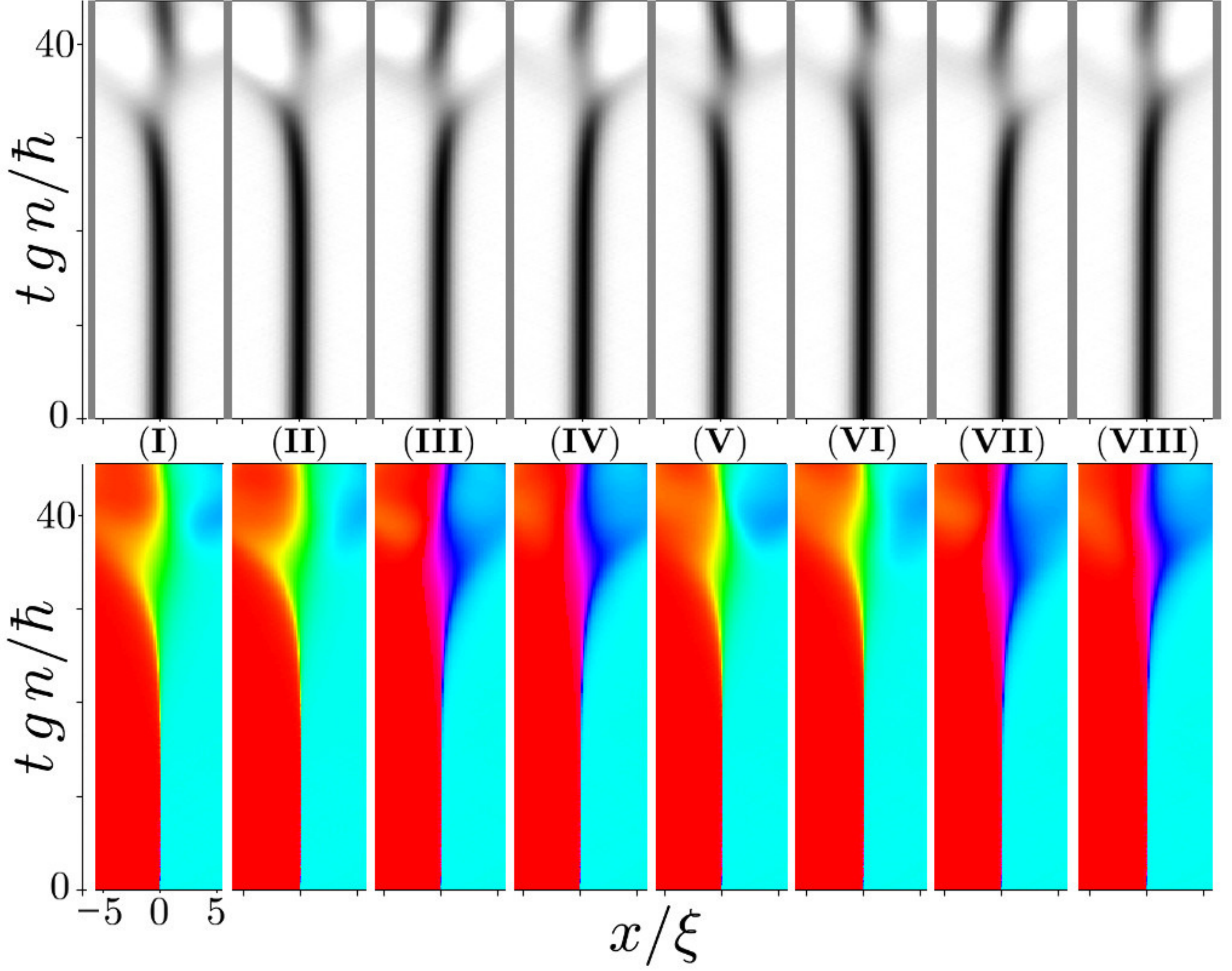}
\caption{Time evolution of an unstable dark soliton stack of
$M=8$  BECs (identified by roman numerals) in a ring configuration with 
linear coupling $\nu/gn = 0.14$ and the same initial conditions as in Fig.\ 
\ref{fig:8comp_stable}. Density (upper panels) and phase (lower panels) are 
shown around the initial soliton position (at $x=0$) as a function of time 
(vertical axis). The dark solitons' decay, consistent with the Bogoliubov 
analysis, is apparent after $t=20$. 
}
\label{fig:8comp_decay}
\end{figure}

In order to observe the dynamical emergence of Josephson vortices, we 
have performed numerical simulations of the time dependent GPE \eqref{eq:tgpe} for the 
real time evolution of overlapping dark solitons in the ring configuration. We have selected a 
study case with a stack of $M=8$ BECs and different values of 
the linear coupling parameter, so that the analytical predictions of the 
stable and unstable regimes of dark solitons can be checked.
A white-noise perturbation, around 1$\%$ of the maximum amplitude, has been 
added to the initial stationary state, as shown in the two left panels of 
Fig.\ \ref{fig:8comp_stable}. Results from two series of simulations are 
reported here, namely for $\nu/gn=0.9$ in the stable regime of the dark 
soliton stack, shown in Fig. 
\ref{fig:8comp_stable} (right panels), and $\nu/gn=0.14$ in the regime of multiple unstable 
modes, depicted in Fig. \ref{fig:8comp_decay}. The results confirm the 
predictions of the linear analysis summarized in Fig.\ \ref{fig:unst_freq}.

In the stable case, the initial configuration survives to the initial weak 
perturbation. The right panels of Fig.\ \ref{fig:8comp_stable} show the time 
evolution of the density and phase profiles, around the soliton position, up to 
$t=60\,\hbar/gn$ for one of the components, the others being indistinguishable 
at the scale of the figure. The situation is quite different for the unstable
case presented in Fig.\ \ref{fig:8comp_decay}, with $\nu/gn=0.14$. In this 
case there are three pairs of unstable modes with momenta $k\xi_y=(\pm1,\pm2, 
\pm3)\times \pi/4$, among which $k\xi_y= \pm2\times \pi/4$ presents the 
maximum unstable frequency $\omega\sim 0.25 \,gn/\hbar$ (see Fig.\ 
\ref{fig:unst_freq}). It is therefore the latter mode that is expected to have 
the highest growth rate during the time evolution, and as a consequence, to 
cause the soliton decay. As can be seen in the graph, the expected 
decay starts between $t=15$ and $20\,\hbar/gn$, a lifetime that depends on 
the type and amount of perturbation. A transient regime follows, between $t=30$ 
and $40\,\hbar/gn$, characterized by wider density depletions and sound wave 
emissions. After this stage, the system reaches a configuration made of four 
Josephson vortices (the associated state bifurcated from the $k\xi_y= 
\pm2\times \pi/4$ modes) in flipped repetition $[ \;\uparrow \uparrow \otimes 
\downarrow \; \downarrow \odot \uparrow \uparrow \otimes 
\downarrow \; \downarrow \odot]$ that does not show further decay. Additionally 
we have checked that this state does not present imaginary frequencies in its 
Bogoliubov spectrum, thus it is dynamically stable.

\section{Discussion and conclusions} \label{sec:concl}

In summary, we have shown that the array of linearly coupled BECs provides an 
excellent playground for the generation and manipulation of Josephson vortices and related solitary waves.
Analytical and numerical evidence was presented for a rich landscape of dynamically stable and unstable configurations. 
In particular the stable configurations should be accessible
for experimental realization with optical lattices, where the 
lattice depth can control the linear coupling between 1D condensates 
\cite{Cazalilla2006,Bloch2008}. Multicomponent condensates with internal 
Josephson coupling (spinor condensates) could also realize a similar situation. The inter-component nonlinearities that are usually present in this case were not considered here and require further analysis along the lines of Ref.~\cite{Sophie2017}. 

Our analysis of the instabilities and decay modes of the dark soliton stack reveals a close similarity to the snaking instability of dark soliton stripes in a continuous, confined superfluid \cite{Brand2002,Mateo2014}. 
There, the system allows for quantized
vibrations of the dark soliton plane according to the linear, normal modes of 
the transverse confinement. The nodal lines from the resulting standing waves 
translate into corresponding patterns of regular vortices after the soliton 
decay. When the transverse dimensions of the system are small compared to its 
axial length, the final stage of this decay leaves the simplest configuration, 
a solitonic vortex \cite{Brand2002}, which is  dynamically stable.
Analogously, in the dark soliton stack, the nodes of the standing wave perturbations across the dark soliton nodes in the BEC array correspond to the cores of the 
emerging Josephson vortices. While decay into a single Josephson vortex is possible in the planar configuration,  the simplest situation in the BEC array in ring configuration is a 
(Josephson) vortex dipole, consisting of a Josephson vortex and a Josephson 
antivortex, which is dynamically stable. In 
addition, for small values of the linear coupling, we have also found 
dynamically stable states made of two Josephson dipoles 
that can be found at the final stage in the decay cascade of dark solitons.

The dark soliton stack of overlapping dark solitons across the array of BECs is
found to be stable under the condition of the inequality \eqref{eq:stabDS}, 
which is a condition on the parameters of nonlinearity $gn$, linear coupling 
$\nu$, and the number of BECs in the discrete array $M$. This gives a number of 
parameters that can be tuned to achieve stability, in particular the linear 
coupling $\nu$, which is controlled by the optical lattice. In contrast, the 
situation of a continuous BEC under trapping with frequency $\omega$ reads 
$\mu/\hbar\omega < 2.65$ \cite{Mateo2014,Muryshev1999}, which is not easily 
reached under typical experimental conditions 
\cite{Becker2013,Ku2014,Donadello2014}.

These facts  suggest a way for the experimental realization of stable dark
solitons in optical lattices. Approximating the stability condition for the dark soliton stack  \eqref{eq:stabDS} for a given number of components $M\gg1$ we obtain
\begin{equation}
% \nu_{1} \approx  \frac{gn}{2({k}\xi_y)^2} = 
\nu \gtrapprox \frac{gn}{2}\left(\frac{M}{2\pi}\right)^2\,.
 \label{eq:nubifap}
\end{equation}
{This condition can be experimentally realized for both the planar and the ring configuration with currently available optical trapping techniques. In the planar case one option is to combine a flat and elongated box potential \cite{Gaunt2013} with a transverse optical lattice. The transverse size of the box defines the number of coupled BEC's, while the height of the optical lattice defines the tunneling coupling $\nu$. The interaction energy is adjustable by either tuning the density of the 1D gas or by exploiting a magnetic Feshbach resonance in order to change $g$. Another option is to populate selected 1D systems in a quantum gas microscope experiment \cite{Sherson2010,Weitenberg}. The special case of $M=2$ can be realized with help of atom chips \cite{Schweigler2017}. The temperature should be chosen low enough to suppress strong thermal fluctuations in the 1D condensates \cite{petrov00}. Quantum fluctuations, which become relevant at low particle number densities, are not expected to destroy the soliton character but may lead to other effects like center-of-mass diffusion \cite{Shamailov2018}.}

{A ring shaped optical lattice as it is required for the ring configuration can be realized by interfering two Laguerre-Gaussian beams with different azimuthal quantum coordinates (i.e.\ topological charges) $l_{1}$ and $l_{2}$ and the same radial quantum coordinate $p=0$ such that only one ring is present in the intensity profile. The number of optical vortices where atoms can be trapped and thus the number of lattice sites is given by the difference in topological charge \cite{Franke-Arnold:07}, $M = \left\vert l_{2}-l_{1}\right\vert$. Laguerre-Gaussian beams can be created from Gaussian laser beams e.g.\ by using spatial light modulators \cite{Franke-Arnold:07} or from circularly polarized light by a combination of an axially symmetric half wave plate and an axially symmetric polarizer \cite{Sakamoto:13}.}

Interesting future extensions of this work on standing waves in a 1D stack of BECs include the characterization of moving solitary waves and a multi-dimensional version where a 1D optical lattice slices a weakly trapped BEC into a stack of 2D pancakes. The results will be presented elsewhere.

\section{Acknowledgements}
CB acknowledges financial support by the DFG within the SFB/TR 49 and by the MAINZ graduate school. Furthermore, CB thanks Massey University for hospitality during a research visit where part of this work was completed. HO acknowledges financial support by the SFB/TR 185 "OSCAR". This work was  supported by the Marsden Fund of New Zealand (Grant No. MAU1604).

%\bibliography{N_components}

%

\end{document}